\documentclass[conference]{IEEEtran}

\usepackage{amsmath}
\usepackage{graphicx}
\usepackage{multirow}
\usepackage{array}
\usepackage{booktabs}
\usepackage[left=54pt, right=54pt, bottom=54pt, top=54pt]{geometry}

\usepackage{url}

\usepackage{xurl}  
\usepackage{hyperref}        
\usepackage[hang,flushmargin]{footmisc}  
\usepackage{cleveref}
\usepackage{etoolbox}
\usepackage{pifont} 
\usepackage{xcolor} 
\usepackage{graphicx} 
\usepackage{makecell}
\usepackage[nolist,nohyperlinks]{acronym}

\definecolor{heavygreen}{RGB}{0,128,0} 

\makeatletter
\patchcmd{\@makecaption}
  {\scshape}
  {}
  {}
  {}
\makeatother

\begin{document}
%
\title{Real-World Deployment of a Lane Change Prediction Architecture Based on Knowledge Graph Embeddings and Bayesian Inference}
%
%
%

\author{M. Manzour$^{1}$, Catherine M. Elias$^{2}$, Omar M. Shehata$^{3}$, R. Izquierdo$^{1}$, and M. A. Sotelo$^{1}$\\
$^{1}$Department of Computer Engineering, University of Alcal\'a, Madrid, Spain\\
$^{2}$ Department of Computer Science, German University in Cairo, Egypt\\
$^{3}$ Department of Mechatronics, German University in Cairo, Egypt\\
$[$ahmed.manzour, ruben.izquierdo, miguel.sotelo$]$@uah.es\\
$[$catherine.elias, omar.shehata$]$@ieee.org
}

\begin{acronym}

\acro{KGE}{Knowledge Graph Embedding}
\acro{CNN}{Convolutional Neural Network}
\acro{RNN}{Recurrent Neural Network}
\acro{LSTM}{Long Short-Term Memory}
\acro{GRU}{Gated Recurrent Unit}
\acro{NGSIM}{Next Generation Simulation}
\acro{CRASH}{CARLA Risky-lane-change Anticipation in Simulated Highways}
\acro{TTC}{Time To Collision}
\acro{THW}{Time Headway}
\acro{RAG}{Retrieval Augmented Generation}
\acro{TCP/IP}{Transmission Control Protocol / Internet Protocol}
\acro{KG}{Knowledge Graph}

\end{acronym}

\maketitle

\begin{abstract}
Research on lane change prediction has gained a lot of momentum in the last couple of years. However, most research is confined to simulation or results obtained from datasets, leaving a gap between algorithmic advances and on-road deployment. This work closes that gap by demonstrating, on real hardware, a lane-change prediction system based on Knowledge Graph Embeddings (KGEs) and Bayesian inference. Moreover, the ego-vehicle employs a longitudinal braking action to ensure the safety of both itself and the surrounding vehicles. Our architecture consists of two modules: (i) a perception module that senses the environment, derives input numerical features, and converts them into linguistic categories; and communicates them to the prediction module; (ii) a pretrained prediction module that executes a KGE and Bayesian inference model to anticipate the target vehicle's maneuver and transforms the prediction into longitudinal braking action. Real-world hardware experimental validation demonstrates that our prediction system anticipates the target vehicle's lane change three to four seconds in advance, providing the ego vehicle sufficient time to react and allowing the target vehicle to make the lane change safely.
\end{abstract}
\begin{IEEEkeywords} 
Lane Change Prediction, Hardware, Experimental Validation, Knowledge Graph Embeddings, Bayesian Inference.
\end{IEEEkeywords}

\IEEEpeerreviewmaketitle

\section{Introduction}
\IEEEPARstart{T}{raffic} accidents are one of the leading causes of death worldwide. As road networks become more complex and the number of vehicles increases, the ability to anticipate the lane change maneuvers of surrounding vehicles becomes not only beneficial but also essential for enhancing road safety. That's why research on lane change prediction has gained significant momentum in recent years. Because when the ego vehicle anticipates the lane change of a surrounding vehicle, it will give the ego vehicle more time to react and create a safe gap for the surrounding vehicle to merge. Most of the research focused on predicting lane changes based on simulations and dataset-driven training approaches. However, these approaches lack real-world validation and may rely on simplified assumptions about perfect sensor accuracy and ideal vehicle behavior. As a result, the effectiveness of these models in real-world scenarios remains uncertain.

This work takes a step toward bridging this gap by presenting a real-world embedded hardware lane change prediction architecture based on \ac{KGE} and Bayesian inference, which anticipates a lane change maneuver and actively makes the ego vehicle respond through a rule-based braking strategy. This not only verifies the prediction quality but also demonstrates that such anticipation can be translated into earlier and enhanced safe reactions on real hardware.
\begin{figure}[t]
\centering
\includegraphics[width=\columnwidth]{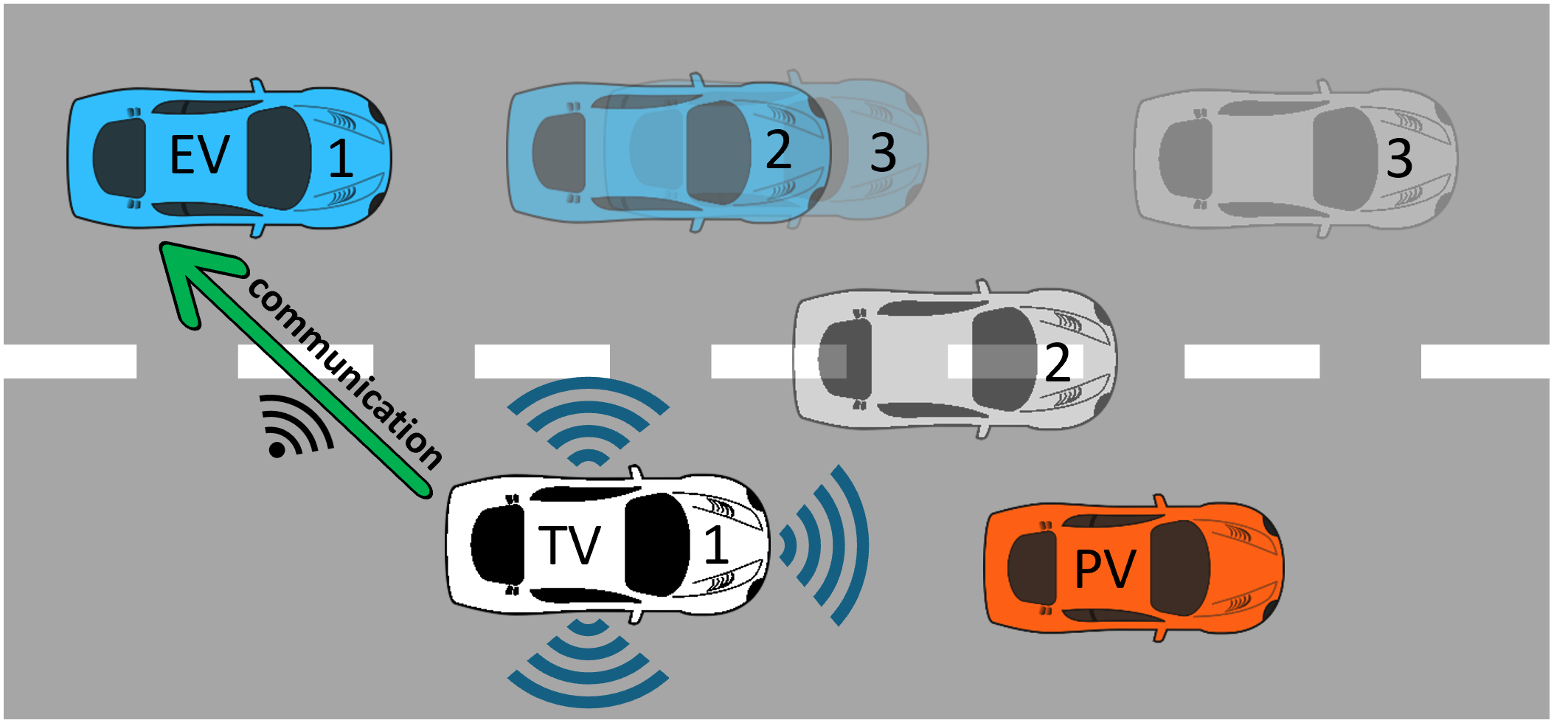}
\caption{The \textit{target vehicle} communicates its sensory data to the \textit{ego vehicle}, allowing the \textit{ego vehicle} to anticipate the \textit{target vehicle}'s maneuver and to brake early and open a safe gap as the \textit{target vehicle} overtakes the slower \textit{preceding vehicle}.}
\label{fig:introduction_lane_changing_example}
\end{figure}
\Cref{fig:introduction_lane_changing_example} illustrates a motivating example for our work, where the target vehicle (TV) is approaching a preceding vehicle (PV), and the ego vehicle (EV) is in the adjacent left lane of the TV, and the TV will make a left lane change. Given our implemented architecture—where the TV communicates sensor data to the EV, and the EV leverages the \ac{KGE} and Bayesian inference model to anticipate lane changes based on this information—the EV can brake early, creating sufficient space for the TV to merge safely. Without this anticipation, the EV would either brake too late or not brake at all, resulting in a high collision risk. This example highlights the real-world importance of implementing a real hardware architecture that has a predictive behavior to prevent accidents before they occur.
After this introduction, the rest of this article is organized as follows. \Cref{sec:sota} presents the state of the art. Then, our proposed architecture is discussed in detail in \cref{sec:methodology}. In \cref{sec:results}, results and discussion will be presented. Finally, \cref{sec:concliusions} concludes the work and provides some recommendations for future work.

\section{State of the Art}\label{sec:sota}
Recently, several studies have focused on predicting vehicle lane changes using different models, including the use of \ac{LSTM} in \cite{han2019driving} and \cite{li2022attention}, a combination of \ac{CNN} and \ac{LSTM} in \cite{izquierdo2019experimental}, \ac{RNN} and \ac{LSTM} in \cite{laimona2020implementation}, and both \ac{LSTM} and \ac{GRU} in \cite{li2021lane}. 
Prediction in \cite{han2019driving} is based on the driver’s characteristics in both lateral and longitudinal driving directions. These characteristics indicate whether the driver has the motive or the incentive to accelerate or decelerate and move in a left or right direction. The prediction involves two main stages: the first extracts the driver’s driving behavior characteristics, and the second stage applies an \ac{LSTM} model to predict the lane change. The model takes inputs representing three seconds of historical data, including longitudinal position, velocity, and lane number for both the target vehicle and its surrounding vehicles.

Works \cite{izquierdo2019experimental} and \cite{laimona2020implementation} both utilize the PREVENTION dataset to extract the target vehicle's features from images. One of the proposed models in \cite{izquierdo2019experimental} is a GoogLeNet-LSTM pipeline. In this setup, the GoogLeNet model is used to extract a feature vector from each image frame, which is then passed along with the target vehicle's bounding box center coordinates and dimensions to an \ac{LSTM} to capture temporal dependencies across a sequence of frames. A comparative analysis between \ac{RNN} and \ac{LSTM} models is conducted by \cite{laimona2020implementation} for the same prediction task. Each model receives a historical sequence of the bounding box centroids of the target vehicle as input, allowing it to learn motion patterns and anticipate lane change behavior.

Lane change and trajectory prediction of a target surrounding vehicle were addressed in \cite{li2021lane} to enable planning a trajectory for the ego vehicle that avoids collisions and ensures safe maneuvers. The pipeline begins by considering the inputs obtained from the \ac{NGSIM} dataset. The input represents the congestion level, which is determined by the longitudinal gap between the target vehicle and surrounding vehicles. A smaller gap means a higher congestion value. This input was fed along with the target vehicle lateral acceleration to the prediction model, which compares two models: the LSTM models and the GRU models. After that, an LQR optimal control model was used to generate candidate trajectories. Then, these candidate trajectories are augmented with the output lane change probabilities from the LSTM model to generate the final predicted trajectory for the target vehicle. Based on this predicted trajectory, a nonlinear predictive control model based on a particle vehicle model is implemented to generate a collision-free trajectory for the ego vehicle. Finally, the motion planning controller is modeled and simulated by Carsim with Simulink. But no hardware experimental validation is considered.

In \cite{li2022attention}, the lane change prediction pipeline is divided into two stages. The first stage is a pre-judgment stage, which states whether the vehicle will make a lane change. It is a binary, discrete value. And this state takes the surrounding vehicle's data, which includes the relative position and velocity between the target vehicle and the surrounding vehicles. This input is fed to a model that combines decision trees and bagging ensemble learning. After getting this pre-judgment value, it is forwarded to the second phase, which is an attention-based LSTM model, in order to predict the final maneuver. It takes a sequence of history from the target vehicle's position, velocity, and acceleration in the longitudinal direction, along with the surrounding vehicle features and the pre-judgment value. The authors mentioned that they need to extend the work by adding more features, enhancing their model, applying it to a more complex road environment, and evaluating it in autonomous driving simulators such as CARLA or Apollo. Finally, the work neither includes real-world testing nor expresses an intention to validate it in real-world conditions. The results are limited to the used dataset, and the future recommendations are confined to validations within simulation environments only.

The works in \cite{manzour2024vehicle} and \cite{manzour2025explainable} focused on predicting lane changes using interpretable and transparent models based on \ac{KGE} and Bayesian inference. This structure enhances interpretability by relying on linguistic inputs, allowing users to better understand the input semantics rather than dealing solely with raw numerical data. The model is also transparent, as it enables users to trace how specific inputs influence the prediction, offering visibility into the reasoning process. This transparency is unlike that of traditional machine learning or deep learning models, which often act as black boxes. In \cite{manzour2024vehicle}, the considered inputs were extracted from the highD dataset \cite{highDdataset} and included the target vehicle’s lateral velocity and acceleration, as well as the \ac{TTC} with the surrounding vehicles. This work focused on predicting only safe lane changes. The work in \cite{manzour2025explainable} extends the scope of \cite{manzour2024vehicle} by addressing both safe and risky lane changes. It achieved this by augmenting the highD dataset (for safe scenarios) with the \ac{CRASH} dataset \cite{manzour2025explainable} (for near-crash and risky maneuvers). In addition to the inputs used in the earlier work, the extended approach also incorporated the target vehicle’s lane ID, its position within the lane, \ac{THW} with the preceding vehicle, the lane with the highest frontal gap, and the lane with the highest attraction score. These features were used to generate a knowledge graph, which was then embedded using the AmbliGraph library \cite{ampligraph}. Bayesian inference was applied on top of these embeddings to compute the final prediction probability. Beyond improving input diversity and prediction scope, \cite{manzour2025explainable} enhanced the model explainability by integrating an \ac{RAG} module, which provides textual explanations for each prediction. Furthermore, it introduced a rule-based longitudinal decision-making mechanism for the ego vehicle, which determines how the vehicle should respond based on the predicted maneuver. The system is evaluated within the CARLA simulation environment.

Despite these advancements, a major limitation remains: none of these systems have been validated in real-world scenarios using embedded hardware. This work addresses that gap by presenting a practical, embedded architecture for lane change prediction, which combines Knowledge Graph Embeddings with Bayesian inference to enable real-time operation on deployed hardware.
\section{Methodology}\label{sec:methodology}
\begin{figure*}[ht]
\centering
\includegraphics[width=\linewidth]{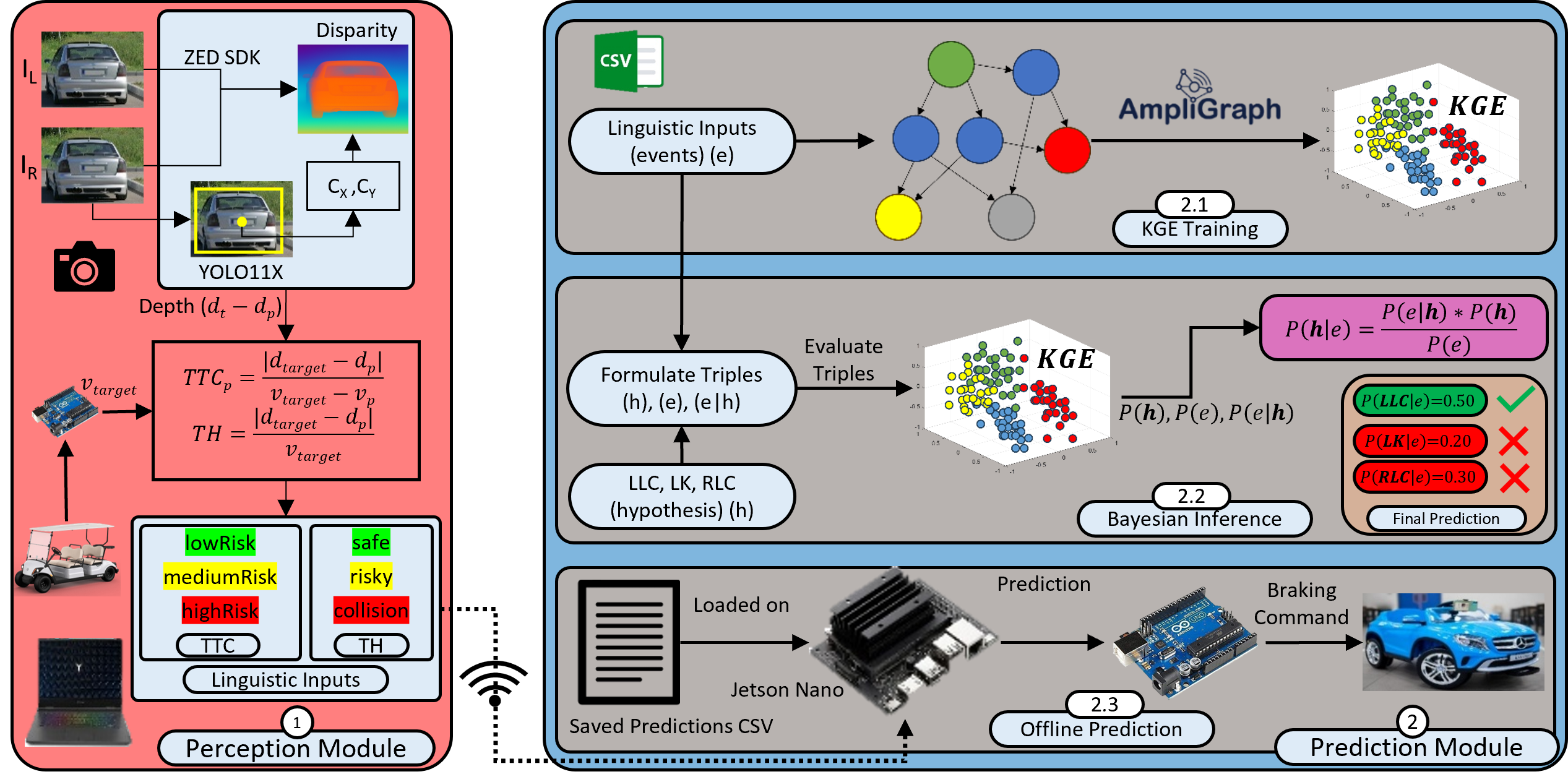}
\caption{System architecture illustrating the hardware-integrated flow of perception, knowledge graph-based prediction, and real-time ego vehicle response.}
\label{fig:methodology}
\end{figure*}

The architecture shown in \Cref{fig:methodology} represents the data flow and component interactions used during the real-world deployment. Originally, our prediction model was designed to process a total of 12 input features during the simulation stage; however, during hardware implementation, only two of them (\ac{TTC} and \ac{THW}) are dynamically computed at runtime by the perception module. The remaining inputs are assigned fixed values because they don’t affect the scenario logic and are constrained by hardware. So, for simplicity, only the \ac{TTC} and \ac{THW} are considered. A more detailed explanation of this feature setup and selection is provided in \Cref{sec:input_config}.
The proposed architecture consists of two modules: perception and prediction. The perception module is loaded on a golf cart, which represents the TV whose upcoming maneuver is to be predicted, and the prediction module is loaded on a blue-scaled that represents the EV, which will brake in case the TV will cut-in its lane. The perception module derives the \ac{TTC} and \ac{THW}, then converts these numerical values into linguistic categories and transmits this data over a \ac{TCP/IP} wireless communication protocol to the prediction module. The prediction module is divided into three stages. The first stage is \ac{KG} training, in which the \ac{KG} model is created and embedded/trained. The second stage uses Bayesian inference as a downstream task on top of the learned embeddings from the \ac{KGE} to generate the prediction. Although both stages are implemented in \cite{manzour2025explainable}, they are briefly outlined here to provide a complete overview of the prediction pipeline. In the third stage, the model's offline version, which contains pre-saved predictions, is exported to a Jetson Nano board \cite{nvidiaJetsonNano2GB} mounted on the EV for real-time validation. The Jetson sends the prediction to a low-level micro-controller via wired serial communication to actuate over the longitudinal control based on the prediction. Each of these functional modules is explained in detail in the following sections.

\subsection{Input Feature Configuration}\label{sec:input_config}
As previously described, the prediction module is originally designed to take 12 inputs, including lateral velocity, lateral acceleration, \ac{TTC} with the left preceding, preceding, right preceding, left following, and right following vehicles, as well as the \ac{THW} with the preceding vehicle, the vehicle’s position inside the lane, lane ID, the lane with the highest attraction score, and the lane with the highest frontal gap. However, due to hardware constraints and to simplify the experimental setup, several assumptions are made: the TV is assumed to be \textit{movingStraight}, with \textit{zeroLateralAcceleration}, and positioned at \textit{centerOfTheLane}. The TV is also assumed to be in the rightmost lane. Given this setup, the \ac{TTC} with the right preceding and following vehicles can be ignored. Additionally, the left lane is assumed to be both free and to offer the highest frontal gap, making it the most attractive for a lane change. Through a focused study on the influence of each input in this two-lane scenario, it was found that the \ac{TTC} with the preceding vehicle has the most significant impact on the prediction output. Even if there is a potential \textit{highRisk} from other vehicles (e.g., a left preceding or following vehicle), a \textit{highRisk} \ac{TTC} with the preceding vehicle will dominate and trigger a lane change to the left. Based on this, the inputs were simplified to focus solely on the \ac{TTC} with the preceding vehicle, and resources are already available to compute the \ac{THW} with the preceding vehicle as well. These two numerical inputs are dynamically computed during runtime. Then, they are converted into linguistic categories and appended to an input list along with the other fixed linguistic categories.

\subsection{Perception Module}
The perception module is responsible for gathering data from sensors and then converting this numerical data into linguistic categories to be fed to the prediction module. The sensor used is the ZED2i stereo camera \cite{stereolabsZED2i}, which provides the depth map. The vehicle is detected and tracked using YOLOv11X, developed by Ultralytics \cite{yolo11_ultralytics}, and the Bitetrack tracker \cite{zhang2022bytetrack}, respectively. The center coordinates \textit{($C_x$, $C_y$)} of the tracked bounding box are then used to obtain the depth, which represents the distance between the TV and the PV. By differentiating this distance over time, the relative velocity between the two vehicles is obtained. The TV’s longitudinal velocity is measured by the micro-controller and transmitted to the computing device. Once the TV’s speed is obtained, it is used along with the depth and relative velocity to compute the \ac{TTC} and \ac{THW} with respect to the PV. These numerical values are then converted into linguistic categories using thresholds that are obtained by following \cite{manzour2025explainable}. For \ac{TTC}, values between $0$ and $4$ seconds are classified as high-risk, between $4$ and $10$ seconds as medium-risk, and values greater than $10$ seconds or any negative value are considered low-risk. For \ac{THW}, values between $0$ and $1$ second indicate a risk of collision (highest risk), values between $1$ and $2$ seconds are considered risky, and values above $2$ seconds are translated to safe. These dynamic features, after being converted into linguistic categories, are appended to a list that also includes the fixed categories described in the previous section. The complete set of linguistic features is then transmitted to the Jetson Nano via \ac{TCP/IP} communication.

\subsection{Prediction Module}
This section provides a brief explanation of the prediction model used. Further details regarding the model architecture, training, and simulation-based testing can be found in \cite{manzour2025explainable}. After receiving the linguistic input categories, they are passed to the prediction system, which outputs whether the vehicle will make a Left Lane Change (LLC), Lane Keep (LK), or Right Lane Change (RLC). This system training process starts with generating the \ac{KG} from our data. A CSV file is used to define the triples \textit{$<$subject, predicate, object$>$} used to construct the graph. For example: \textit{$<$vehicle, HAS\_CHILD, vehicle1$>$}, \textit{$<$vehicle1, PRECEDING\_VEHICLE\_TTC\_IS, highRiskPreceding$>$},\textit{$<$vehicle1, INTENTION\_IS, LLC$>$}. In this way, the \ac{KG} is structured. Once the graph is built, it is embedded using the Ampligraph library. Embedding refers to the process of converting each node and relation into vector representations, allowing us to use them for various downstream tasks, such as machine learning and Bayesian inference. Nodes that are connected by a relation end up with nearby embeddings, while unrelated nodes are placed far from each other in the embedding space.







After the training phase, Bayesian inference is performed as a downstream task on top of the learned embeddings. The hypotheses are defined as: LLC, LK, and RLC. Events derived from sensory data are also available, having been previously converted into linguistic categories. Both the hypotheses and the events are written as triples, for instance, \textit{$<$vehicle, INTENTION\_IS, LLC$>$} for a hypothesis (h), and \textit{$<$vehicle, PRECEDING\_VEHICLE\_TTC\_IS, highRiskPreceding$>$} for an event (e). Then, the triples of the event given the hypothesis (e$|$h) is formulated, like \textit{$<$highRiskPreceding, INTENTION\_IS, LLC$>$}, which asks, what is the probability that the \textit{highRiskPreceding} event is true given that the \textit{LLC} hypothesis is true. The \ac{KGE} model then evaluates these triples and returns a probability score that indicates how likely each triple is true. Then, the three probabilities (P(h), P(e$|$h), and P(e)) are fed to \Cref{eq:bayesian} to get P(h$|$e). This is done for each hypothesis (LLC, LK, RLC), and the final scores are compared. The one with the highest score is selected as the predicted intention.
\begin{equation}
    P(h|e)=\frac{P(h)P(e|h)}{P(e)}
    \label{eq:bayesian}
\end{equation}

To enable deployment on the Jetson Nano board, predictions are pre-computed for every possible linguistic input combination and stored in a CSV file, taking advantage of the structured nature of the linguistic inputs. Later, when a new input is received during runtime, the system simply searches for the matching linguistic input in the CSV and fetches the corresponding prediction. This is one of the key advantages of working with linguistic categories instead of raw numerical data; it allows us to discretize and reduce the input space and easily map inputs to predictions. So, briefly, the prediction module on the Jetson Nano takes the input as a list of linguistic categories, searches the CSV for a match, retrieves the predicted intention, and sends it to the micro-controller using wired serial communication. The braking logic of the EV is based on the communicated prediction: if the predicted maneuver is an LLC, the vehicle stops; if it is an LK, the vehicle continues moving at its designated speed.

\section{Results}\label{sec:results}
\begin{figure}[ht]
\centering
\includegraphics[width=\columnwidth]{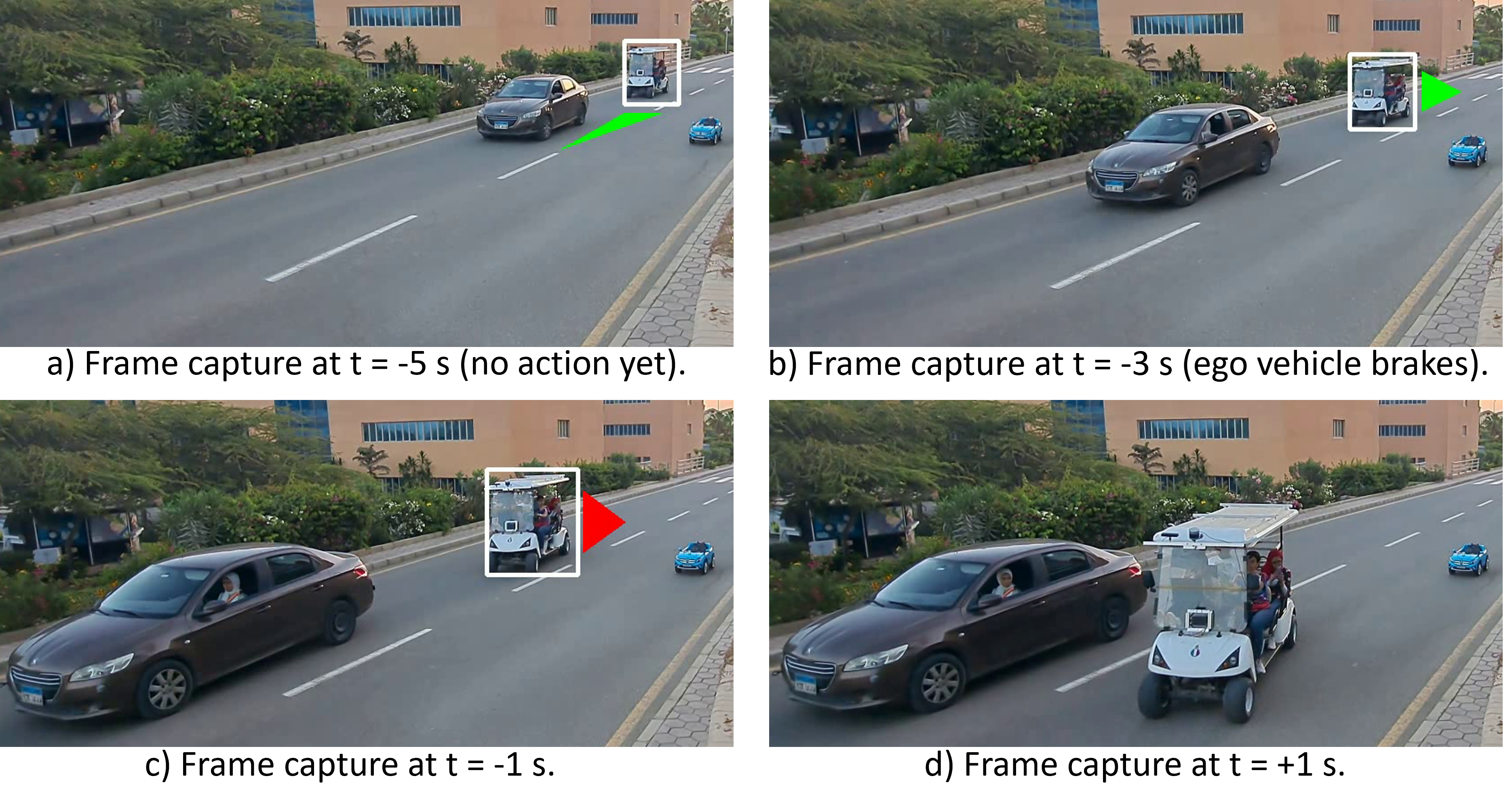}
\caption{Validating a hardware scenario using the proposed prediction model. In 3a (five seconds before crossing the lane marking), the target vehicle is moving straight, and the model predicts a lane-keeping maneuver. Then, the model anticipates a safe left lane change as in 3b, prompting the ego vehicle to stop. After that, in 3c, the model predicts a risky left lane change, and the ego vehicle has already come to a stop, allowing the target vehicle to complete the lane change, as shown in 3d.
}
\label{fig:with_prediction_scenario}
\end{figure}

\begin{figure}[ht]
\centering
\includegraphics[width=\columnwidth]{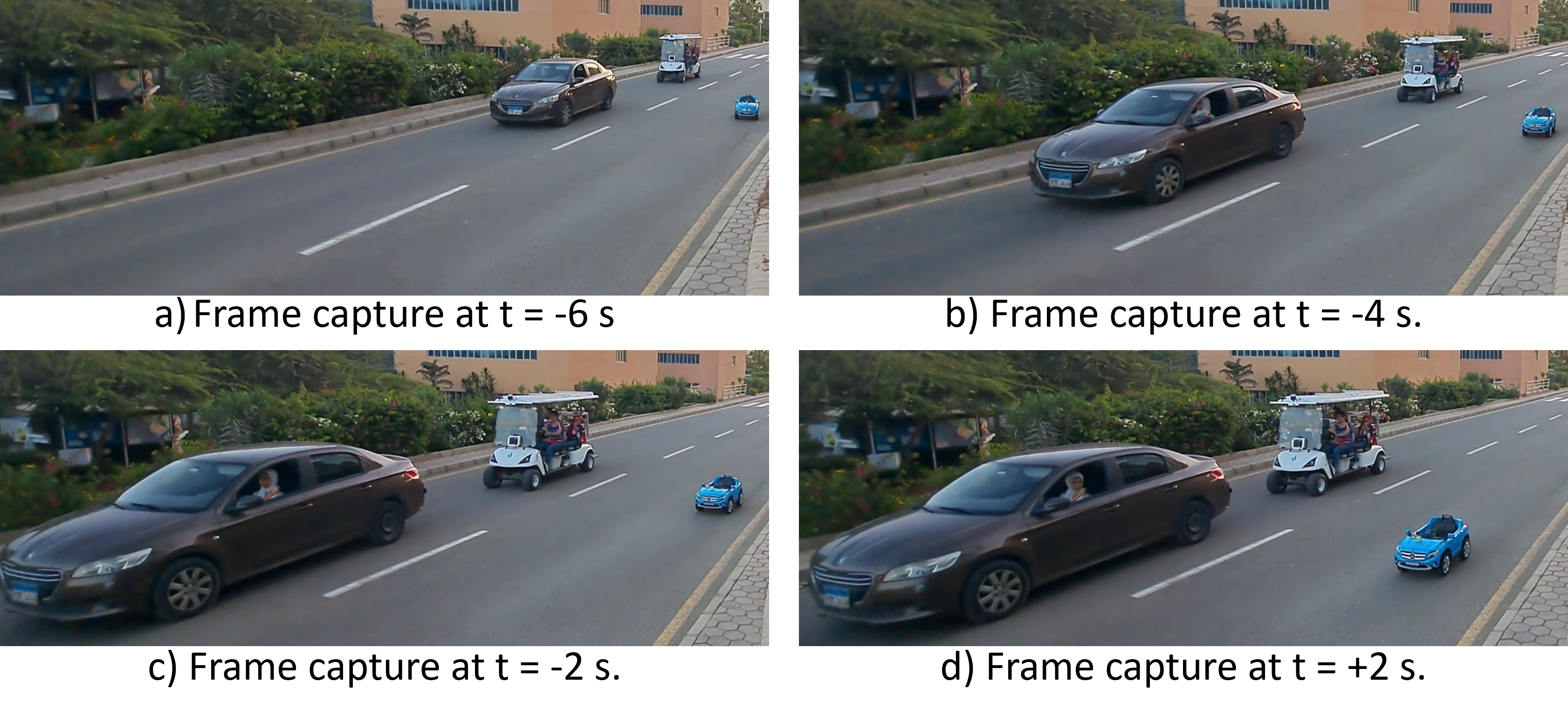}
\caption{Demonstration of the ego vehicle’s behavior without using the prediction model. In 4a (six seconds before the target vehicle comes to a complete stop), both the target vehicle and the ego vehicle are moving normally. In 4b and 4c, the ego vehicle fails to anticipate the target vehicle’s lane change maneuver due to the absence of prediction. As a result, no braking action is taken by the ego vehicle, leaving no opportunity for the target vehicle to safely change lanes. This forces the target vehicle to brake aggressively to avoid a collision with the preceding vehicle.}
\label{fig:without_prediction_scenario}
\end{figure}

The evaluation of this experiment involves two scenarios that follow the same setup. The only difference between them is that, in one scenario, our prediction model is installed on the EV, while in the other, it is not. This means that no prediction is carried out in the second scenario. Both scenarios begin with the same initial positions for all vehicles and speed profiles until the interaction. These two scenarios are illustrated in \Cref{fig:with_prediction_scenario} and \Cref{fig:without_prediction_scenario}.
The road has two lanes: the EV is placed in the left lane, while the TV and the PV are in the right lane.  After the EV starts to move, the TV and the PV start to move with the same velocity until the PV reaches a certain point and begins to brake. This triggers the need for the TV to change to the left lane. Here is where the presence or absence of the prediction model makes the difference. When the model is active, the EV anticipates the lane change and starts braking earlier, as in \Cref{fig:with_prediction_scenario}b. This early response opens a gap, allowing the TV to merge safely. On the other hand, when the model is not active, the EV maintains its speed and does not react. As a result, it does not leave enough space for the TV to change lanes, as shown in \Cref{fig:without_prediction_scenario}c. This forces the TV to brake hard to avoid a collision with the PV and EV.

\begin{figure}[ht]
\centering
\includegraphics[width=\columnwidth]{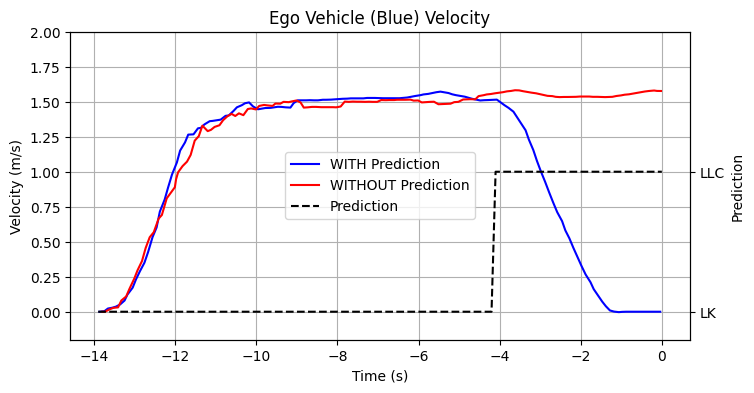}
\caption{Comparison of the ego vehicle’s velocity with and without prediction. The ego vehicle slows down earlier when prediction is integrated, providing more time for the target vehicle to complete the lane change. The moment labeled as t = 0 seconds corresponds to the time at which the target vehicle crosses the lane marking. Negative time values indicate moments before the lane-changing event, and positive time values represent moments after the lane change. }
\label{fig:ego_vehicle_velocity}
\end{figure}

\begin{figure}[ht]
\centering
\includegraphics[width=\columnwidth]{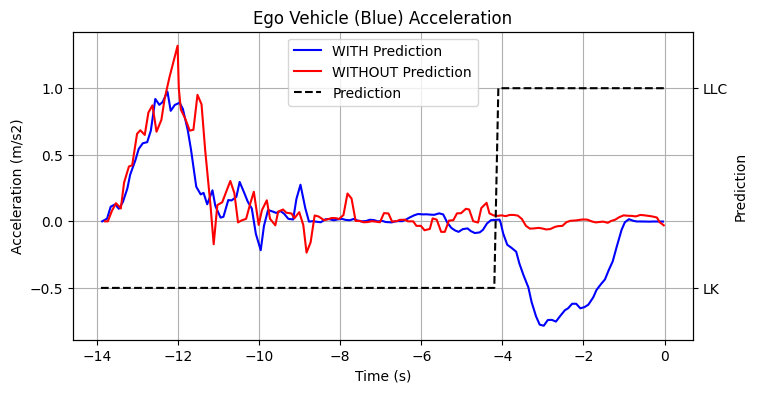}
\caption{Comparison of the ego vehicle's acceleration with and without prediction, showing early deceleration when lane change is anticipated.}
\label{fig:ego_vehicle_acceleration}
\end{figure}

To evaluate the impact of both scenarios, the velocity and acceleration curves of the EV and TV are examined. Starting with the EV: in the first scenario, where the prediction model is installed, the velocity curve in \Cref{fig:ego_vehicle_velocity} shows an early decrease, indicating that the vehicle started braking after predicting the upcoming left lane change $4$ seconds in advance. In the second scenario, where no model is present, the EV’s velocity is constant, as there is no anticipation of the lane change. Looking at the acceleration profile in \Cref{fig:ego_vehicle_acceleration}, in scenario $1$, the EV clearly begins to decelerate, and as it comes to a stop, the acceleration gradually returns to zero. In contrast, in scenario $2$, the acceleration curve stays close to zero, indicating motion with constant speed.
\begin{figure}[ht]
\centering
\includegraphics[width=\columnwidth]{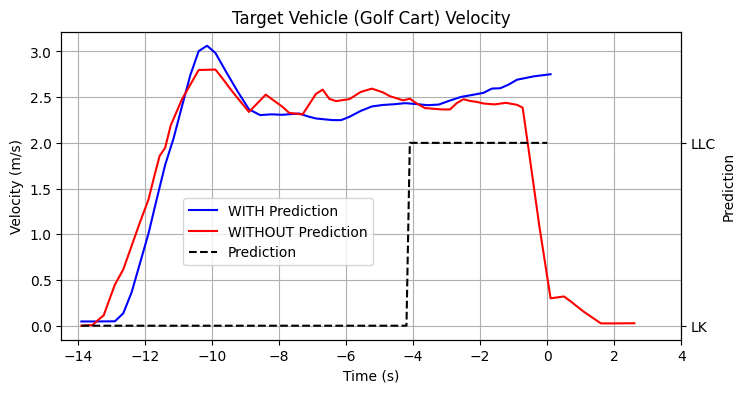}
\caption{Comparison of the target vehicle’s velocity with and without prediction. Velocity of the target vehicle with and without integrated prediction on the ego vehicle. Without prediction, the target vehicle decelerates sharply. With prediction, the ego vehicle yields earlier, allowing the target to maintain a smoother trajectory.}
\label{fig:target_vehicle_velocity}
\end{figure}
\begin{figure}[ht]
\centering
\includegraphics[width=\columnwidth]{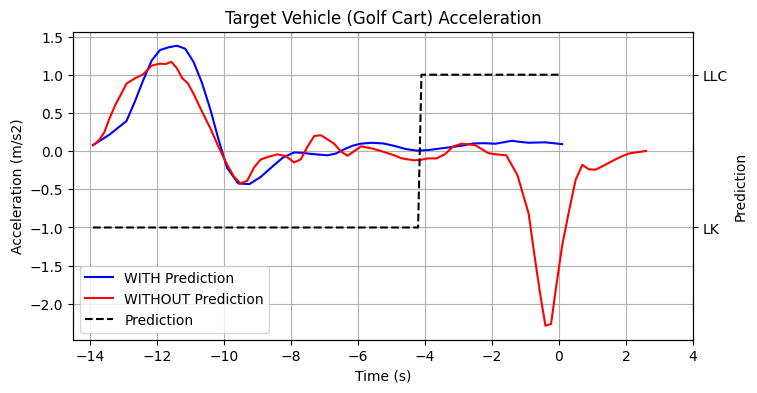}
\caption{Comparison of acceleration profile of the target vehicle (golf cart) when the ego vehicle has prediction enabled versus disabled. Without prediction, the target executes abrupt braking due to the ego vehicle not yielding.}
\label{fig:target_vehicle_acceleration}
\end{figure}

For the TV, the difference in behavior is also clear. In scenario $1$, its longitudinal velocity remains stable as shown in \Cref{fig:target_vehicle_velocity}.
This is because the EV anticipated the lane change and made space for the TV to maneuver. However, in scenario $2$, the TV’s velocity shows a sudden drop. That drop is caused by the aggressive braking required when the EV did not slow down or leave space. This difference becomes even more obvious when looking at the acceleration curves in \Cref{fig:target_vehicle_acceleration}.
In scenario 1, the TV continues forward steadily. But in scenario 2, there’s a sharp drop in acceleration, highlighting the sudden braking. This kind of maneuver introduces risks and discomfort for the passengers. Additionally, the implemented prediction architecture achieves a computational performance of approximately $15$ iterations per second, confirming its capability for real-time prediction and decision-making scenarios. Finally, for better visualization, \autoref{tab:media} includes links to videos that provide results from two different views for each scenario. More views can be accessed from the following playlist link: \url{https://www.youtube.com/playlist?list=PLAeK3AuwxenFqIeAnKa9BLB8bDqEk8dNH}
\begin{table}[ht]
\caption{Video links showing hardware validation results from two different viewpoints, with and without integrating the prediction model.}
\begin{center}
\begin{tabular}{|c|c|}
\hline
      Scenario View          & Link\\
\hline
WITH Prediction (View 1)   & \url{https://youtu.be/p4LKoaUsZpc}\\
\hline
WITHOUT Prediction (View 1)   & \url{https://youtu.be/BFVqhoT5Lck}\\
\hline
WITH Prediction (View 2)   & \url{https://youtu.be/8Dc6Agxgn8M}\\
\hline
WITHOUT Prediction (View 2)   & \url{https://youtu.be/U7PYoEeVnqY}\\

\hline
\end{tabular}
\label{tab:media}
\end{center}
\end{table}

\section{Conclusion}\label{sec:concliusions}
This work presents a real hardware implementation for predicting lane changes. The system consists of two modules. The first module is the perception, which collects numerical data from sensors, converts it into linguistic categories, and transmits the linguistic data to the prediction module via a TCP/IP communication protocol. The second module is the prediction, which receives the linguistic inputs and anticipates the lane change behavior of the TV. This module is built using \ac{KGE} combined with Bayesian inference. An offline version of the prediction model, containing precomputed predictions, is deployed on a Jetson Nano. The predicted maneuver is sent to a low-level controller, which determines whether the EV should stop or continue moving based on the received prediction. Our system is capable of predicting the TV's lane change $3\sim4$ seconds in advance, allowing the EV sufficient time to respond and enabling the TV to merge safely. Future work can focus on considering more inputs for broader scenarios (e.g, two and three-lane roads), investigating other methods and hardware components in the perception and prediction modules, and decoupling direct communication between the TV and EV by introducing an intermediate relay module. This relay would manage data transmission and additional computations, allowing the perception module to focus solely on processing sensor data.
\section*{Acknowledgment}
This research has been funded by the A-IQ-Ready project of the
KDT Program of the European Commission under Grant Agreement: 101096658.

\ifCLASSOPTIONcaptionsoff
  \newpage
\fi



%

\bibliographystyle{IEEEtran}
\bibliography{IEEEabrv,bibtex/bib/IEEEexample}

\begin{thebibliography}{10}
\providecommand{\url}[1]{#1}
\csname url@samestyle\endcsname
\providecommand{\newblock}{\relax}
\providecommand{\bibinfo}[2]{#2}
\providecommand{\BIBentrySTDinterwordspacing}{\spaceskip=0pt\relax}
\providecommand{\BIBentryALTinterwordstretchfactor}{4}
\providecommand{\BIBentryALTinterwordspacing}{\spaceskip=\fontdimen2\font plus
\BIBentryALTinterwordstretchfactor\fontdimen3\font minus \fontdimen4\font\relax}
\providecommand{\BIBforeignlanguage}[2]{{%
\expandafter\ifx\csname l@#1\endcsname\relax
\typeout{** WARNING: IEEEtran.bst: No hyphenation pattern has been}%
\typeout{** loaded for the language `#1'. Using the pattern for}%
\typeout{** the default language instead.}%
\else
\language=\csname l@#1\endcsname
\fi
#2}}
\providecommand{\BIBdecl}{\relax}
\BIBdecl

\bibitem{han2019driving}
T.~Han, J.~Jing, and {\"U}.~{\"O}zg{\"u}ner, ``Driving intention recognition and lane change prediction on the highway,'' in \emph{2019 IEEE Intelligent Vehicles Symposium (IV)}.\hskip 1em plus 0.5em minus 0.4em\relax IEEE, 2019, pp. 957--962.

\bibitem{li2022attention}
Z.-N. Li, X.-H. Huang, T.~Mu, and J.~Wang, ``Attention-based lane change and crash risk prediction model in highways,'' \emph{IEEE transactions on intelligent transportation systems}, vol.~23, no.~12, pp. 22\,909--22\,922, 2022.

\bibitem{izquierdo2019experimental}
R.~Izquierdo, A.~Quintanar, I.~Parra, D.~Fern{\'a}ndez-Llorca, and M.~A. Sotelo, ``Experimental validation of lane-change intention prediction methodologies based on cnn and lstm,'' in \emph{2019 IEEE Intelligent Transportation Systems Conference (ITSC)}.\hskip 1em plus 0.5em minus 0.4em\relax IEEE, 2019, pp. 3657--3662.

\bibitem{laimona2020implementation}
O.~Laimona, M.~A. Manzour, O.~M. Shehata, and E.~I. Morgan, ``Implementation and evaluation of an enhanced intention prediction algorithm for lane-changing scenarios on highway roads,'' in \emph{2020 2nd Novel Intelligent and Leading Emerging Sciences Conference (NILES)}.\hskip 1em plus 0.5em minus 0.4em\relax IEEE, 2020, pp. 128--133.

\bibitem{li2021lane}
L.~Li, W.~Zhao, C.~Xu, C.~Wang, Q.~Chen, and S.~Dai, ``Lane-change intention inference based on rnn for autonomous driving on highways,'' \emph{IEEE Transactions on Vehicular Technology}, vol.~70, no.~6, pp. 5499--5510, 2021.

\bibitem{manzour2024vehicle}
M.~Manzour, A.~Ballardini, R.~Izquierdo, and M.~Sotelo, ``Vehicle lane change prediction based on knowledge graph embeddings and bayesian inference,'' in \emph{2024 IEEE Intelligent Vehicles Symposium (IV)}.\hskip 1em plus 0.5em minus 0.4em\relax IEEE, 2024, pp. 1893--1900.

\bibitem{manzour2025explainable}
\BIBentryALTinterwordspacing
M.~Manzour, A.~Ballardini, R.~Izquierdo, and M.~{\'A}. Sotelo, ``Explainable lane change prediction for near-crash scenarios using knowledge graph embeddings and retrieval augmented generation,'' 2025. [Online]. Available: \url{https://arxiv.org/abs/2501.11560}
\BIBentrySTDinterwordspacing

\bibitem{highDdataset}
R.~Krajewski, J.~Bock, L.~Kloeker, and L.~Eckstein, ``The highd dataset: A drone dataset of naturalistic vehicle trajectories on german highways for validation of highly automated driving systems,'' in \emph{2018 21st International Conference on Intelligent Transportation Systems (ITSC)}, 2018, pp. 2118--2125.

\bibitem{ampligraph}
\BIBentryALTinterwordspacing
L.~Costabello, S.~Pai, C.~L. Van, R.~McGrath, N.~McCarthy, and P.~Tabacof, ``{AmpliGraph: a Library for Representation Learning on Knowledge Graphs},'' Mar. 2019. [Online]. Available: \url{https://doi.org/10.5281/zenodo.2595043}
\BIBentrySTDinterwordspacing

\bibitem{nvidiaJetsonNano2GB}
N.~Corporation, ``Jetson nano 2gb developer kit,'' \url{https://developer.nvidia.com/embedded/learn/get-started-jetson-nano-2gb-devkit}, 2020, accessed: 2025-06-11.

\bibitem{stereolabsZED2i}
S.~Inc., ``{ZED 2i Stereo Camera},'' \url{https://www.stereolabs.com/store/products/zed-2i}, 2025, accessed: 2025-06-11.

\bibitem{yolo11_ultralytics}
\BIBentryALTinterwordspacing
G.~Jocher and J.~Qiu, ``Ultralytics yolo11,'' 2024. [Online]. Available: \url{https://github.com/ultralytics/ultralytics}
\BIBentrySTDinterwordspacing

\bibitem{zhang2022bytetrack}
Y.~Zhang, P.~Sun, Y.~Jiang, D.~Yu, F.~Weng, Z.~Yuan, P.~Luo, W.~Liu, and X.~Wang, ``Bytetrack: Multi-object tracking by associating every detection box,'' in \emph{European conference on computer vision}.\hskip 1em plus 0.5em minus 0.4em\relax Springer, 2022, pp. 1--21.

\end{thebibliography}

%








\end{document}